# Designing a WISHBONE Protocol Network Adapter for an Asynchronous Network-on-Chip

Ahmed H.M. Soliman[1], E.M. Saad[2], M. El-Bably[3] and Hesham M. A. M. Keshk[4]

**Electronics, Communications, and Computer Engineering Department, University of Helwan**
**Cairo, Egypt 11795, Helwan**

**Abstract**

The Scaling of microchip technologies, from micron to sub-micron and now to deep sub-micron (DSM) range, has enabled large scale systems-on-chip (SoC). In future deep submicron (DSM) designs, the interconnect effect will definitely dominate performance. Network-on-Chip (NoC) has become a promising solution to bus-based communication infrastructure limitations. NoC designs usually targets Application Specific Integrated Circuits (ASICs), however, the fabrication process costs a lot. Implementing a NoC on an FPGA does not only reduce the cost but also decreases programming and verification cycles. In this paper, an Asynchronous NoC has been implemented on a SPARTAN-3E® device. The NoC supports basic transactions of both widely used on-chip interconnection standards, the Open Core Protocol (OCP) and the WISHBONE Protocol. Although, FPGA devices are synchronous in nature, it has been shown that they can be used to prototype a Global Asynchronous Local Synchronous (GALS) systems, comprising an Asynchronous NoC connecting IP cores operating in different clock domains.

## 1. Introduction

In future deep submicron (DSM) designs, the interconnect effect will definitely dominate performance [1], and Networks-on-Chip has become a promising solution to present bus-based communication infrastructure limitations [2]. Usually NoC area dominates the target SoC devices and thus greatly influences power consumption and increases the clock distribution/skew problem. Asynchronous designs have merits over familiar synchronous circuit designs. These merits have been highlighted recently. The low power consumption, high operating speed and low electromagnetic noise of asynchronous circuits have attracted designers to learn more about asynchronous circuit designs. Some useful materials in the literature have been found to cover this field especially [3].

NoC designs usually targets Application Specific Integrated Circuits (ASICs), however, the fabrication process costs a lot. Implementing a NoC on an FPGA does not only reduce the cost but also decreases programming and verification cycles. FPGA devices are primarily being built in order to support synchronous digital designs. Montage was the first attempt to explicitly support asynchronous design realizations on FPGA devices [4]. S. C. Smith has proposed a design of a logic element for implementing an asynchronous FPGA [5]. However, ordinary FPGA devices can realize Asynchronous circuits provided that the designer should carry out an intensive timing analysis. A design methodology for implementing asynchronous circuits on an FPGA has been formulated by J. N. Lassen [6]. Thang has introduced a basic NoC architecture (BASIC-NoC) which has been successfully prototyped on a SPARTAN-3E device [2]. However, it is a synchronous NoC realization. J.N. Lassen has designed and prototyped an asynchronous NoC on an FPGA device [6], However, the network adapters for this NoC only support the Open Core Protocol (OCP) [7] and does not support other protocol standards such as WISHBONE protocol [8], and hence, this NoC cannot be directly used to interconnect Intellectual Property (IP) cores supporting other standards.

In this paper, the Asynchronous NoC developed by [6] has been extended to support WISHBONE Interconnection architecture [8]. This paper is organized as follows: an overview of the Asynchronous NoC being used is described in section 2. Router architecture is shown in section 3. In Section 4, an overview of the modified network adapters developed in this work is explained. Section 5 presents simulation and synthesis results while Section 6 concludes the paper.

## 2. Asynchronous NoC

NoC or generally speaking any network design can be characterized by at least three main parameters: Topology, Routing Algorithm, and Flow Control Mechanism. Fairly complex algorithms provide better performance; however, these algorithms are sometimes infeasible to be implemented on NoCs due to their resource constrained nature [9]. This is especially true for FPGA devices. The asynchronous NoC design that has been used in this work is adopted from [6].

### 2.1 Topology

Mesh topology has been selected due to its layout efficiency, ease of implementation, and scalability. Although the topology is mesh from the start of design however it can be easily converted into torus. Figure1 shows a layout for the NoC Architecture.

### 2.2 Routing Algorithm

Routing algorithms can be classified into deterministic routing, partially adaptive routing, and fully adaptive routing. Although adaptive algorithms results in improved network bandwidth and throughput, however, these algorithms are not efficient in terms of chip area, power consumption and processing overhead. Thus, in a resource constrained environment deterministic routing is preferred. In this NoC, deterministic source routing has been employed. The destination address is supplied by the source in the header of the packet, where two bits are allocated in order to encode the requested output port direction at each router. It is worth mentioning that the inner router nodes have five ports, four compass ports (North, East, South, and West) and one local port. The local output port is encoded by using the same port address from which the packet has arrived to the router.

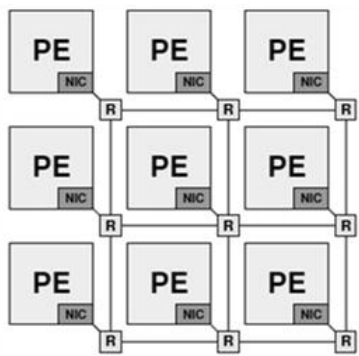

Fig. 1 NoC Layout.

### 2.3 Flow Control Mechanism

In order to simplify the design further a flow control is suggested to operate on a higher level. End-to-end flow control is being employed. Routers in this case have no role in flow control hence if a p acket suffers from congestion it is simply dropped. It is the responsibility of the communicating Processing Elements (PE) s to negotiate packet successful arrival through special flow control packets.

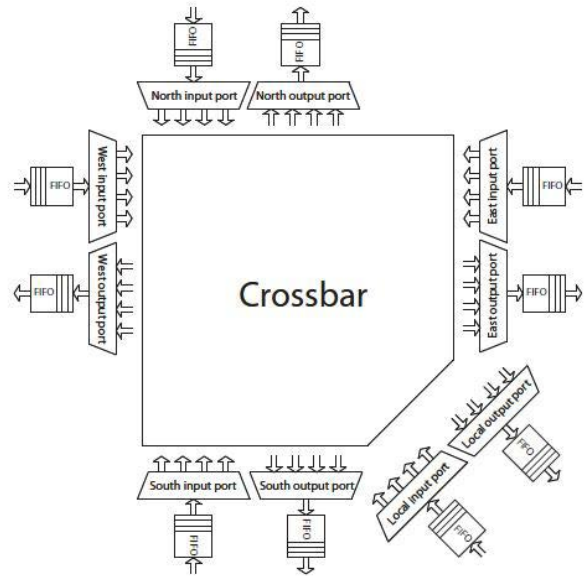

Fig. 2 Router Architecture

## 3. Router Architecture

The developed simple NoC router architecture primarily consists of the following main components: mutex, crossbar, input ports, output ports, merge, and FIFO buffers. The router architecture is shown in Figure 2. Mutex is used for arbitration, while the input port extracts and decodes the destination address from the header flit while, merge components are used to deliver flits through output ports. VHDL language was used by [6] to specify the router internal design in a structural manner.

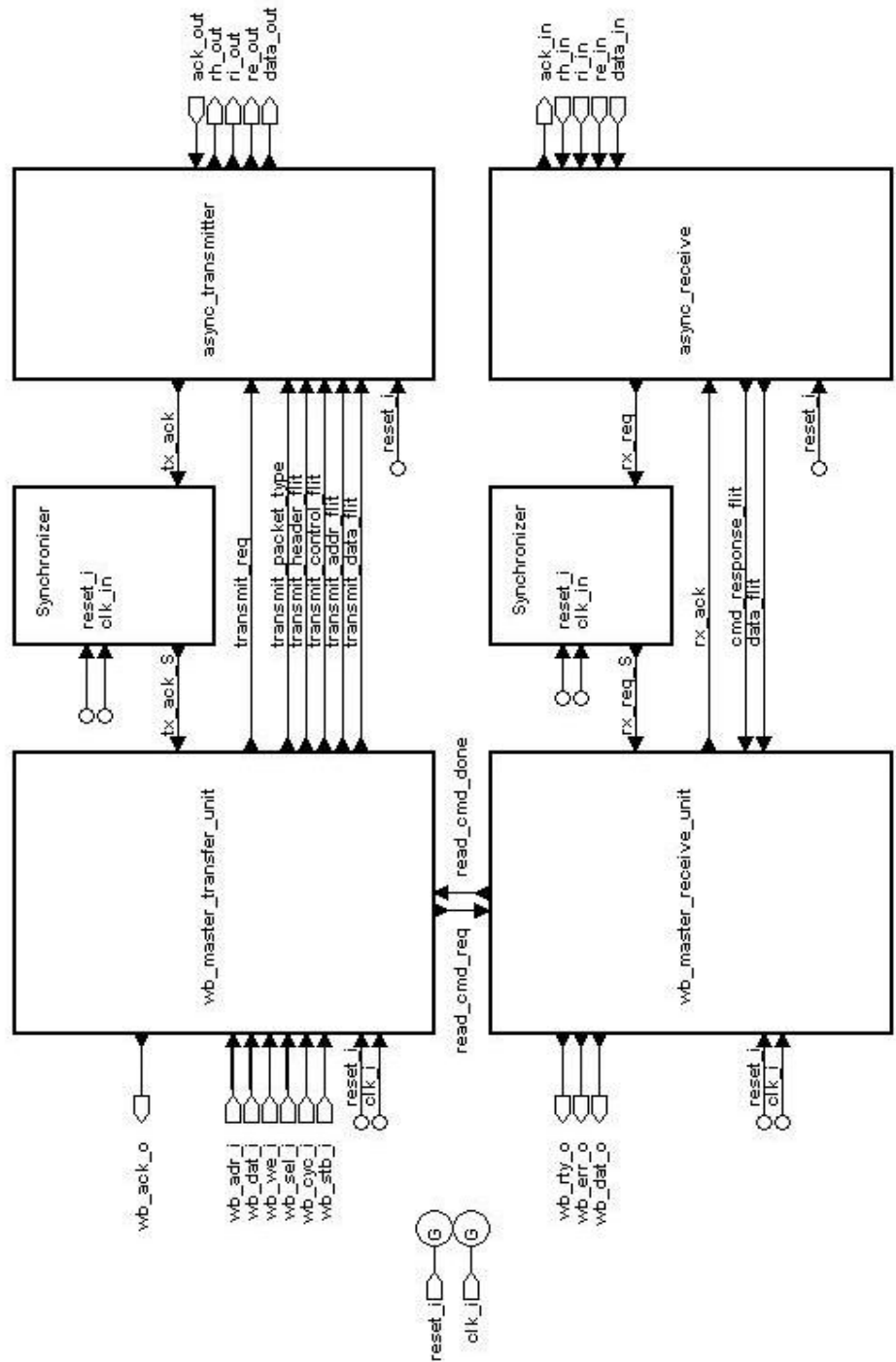

Fig. 3 WISHBONE Master NA.

## 4. Network Adapters

In order to exploit any NoC within a SoC design, as well as increasing its re-usability, network adapters that support different on-chip protocols have to be designed. This enables the designer to easily plug standard IP cores. Since the WISHBONE interconnect protocol is one of the most popular standards adopted by IP designers and as stated previously the Network adapters designed in [6] supports the OCP standard protocol only, thus, in this paper a Network Adapter was designed in order to add a new feature to this NoC. The new added feature is to support the WISHBONE protocol standard.

The network adapter (NA) consists of two interfaces. A Core Interface (CI) and a Network Interface (NI). Also there are two types of Network Adapters. A master (Slave) NA is used to connect a master (Slave) IP core.

4.1 Modified Master Network Adapters

The master network adapter consists of the following components: Core Interface (master transmit unit and master receive unit), Network Interface (Asynchronous transmitter and Asynchronous receiver), and two Synchronizer units which are used at the boundaries between synchronous and asynchronous domains. Figure 3 displays the block diagram for the WISHBONE master NA.

Figure 4 shows the ASM chart for the WISHBONE transfer unit. The controller for this unit remains on the wait state until a WISHBONE cycle starts, and a strobe signal was asserted high. Afterwards, the controller performs state transition to the store packet state. in this state all WISHBONE signals are stored temporarily on internal buffers. After storing WISHBONE signals the controller transfers to the route look-up state, in which the address look-up table was accessed in order to extract the slave node routing address in the NoC, this address was found from the highest 4-bits in the 32-bits WISHBONE address for the slave target. The route look-up process requires a complete clock cycle, afterwards, the controller starts handshaking with the asynchronous transmitter by asserting the request signal through the (Req) state. The controller remains in that state until acknowledge signal was received from the asynchronous transmitter, then the controller starts asserting its acknowledge signal through the (Ack) state and don't negate it until the asynchronous transmitter acknowledge was negated, then, the controller returns back to the wait state waiting for another WISHBONE transaction to begin.

Figure 5 shows the ASM chart for the WISHBONE receive unit.

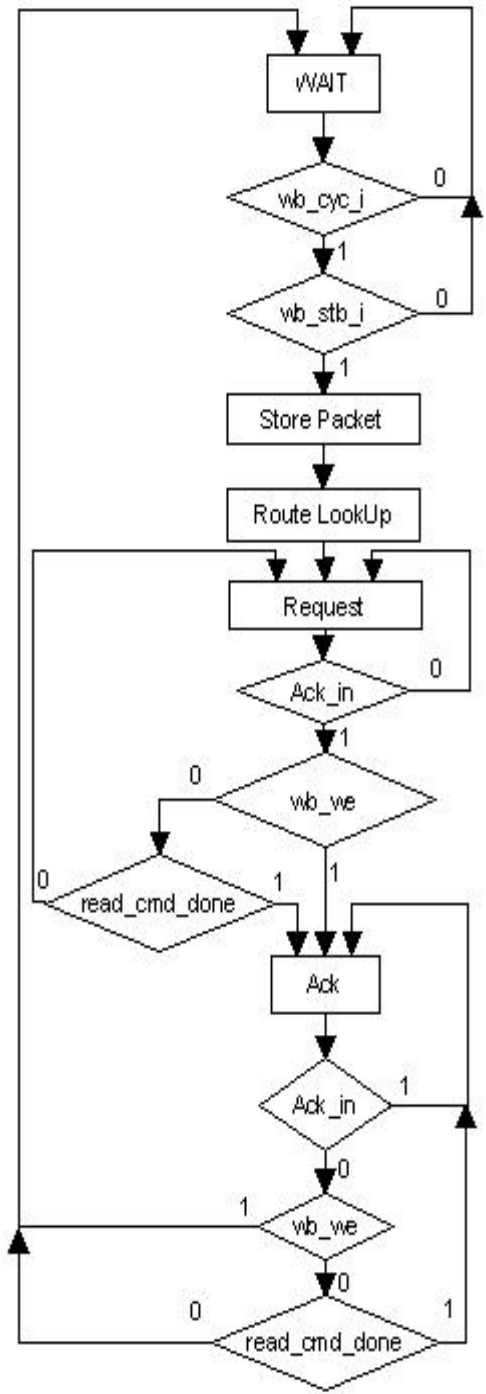

Fig. 4 ASM chart for the WISHBONE master transfer unit.

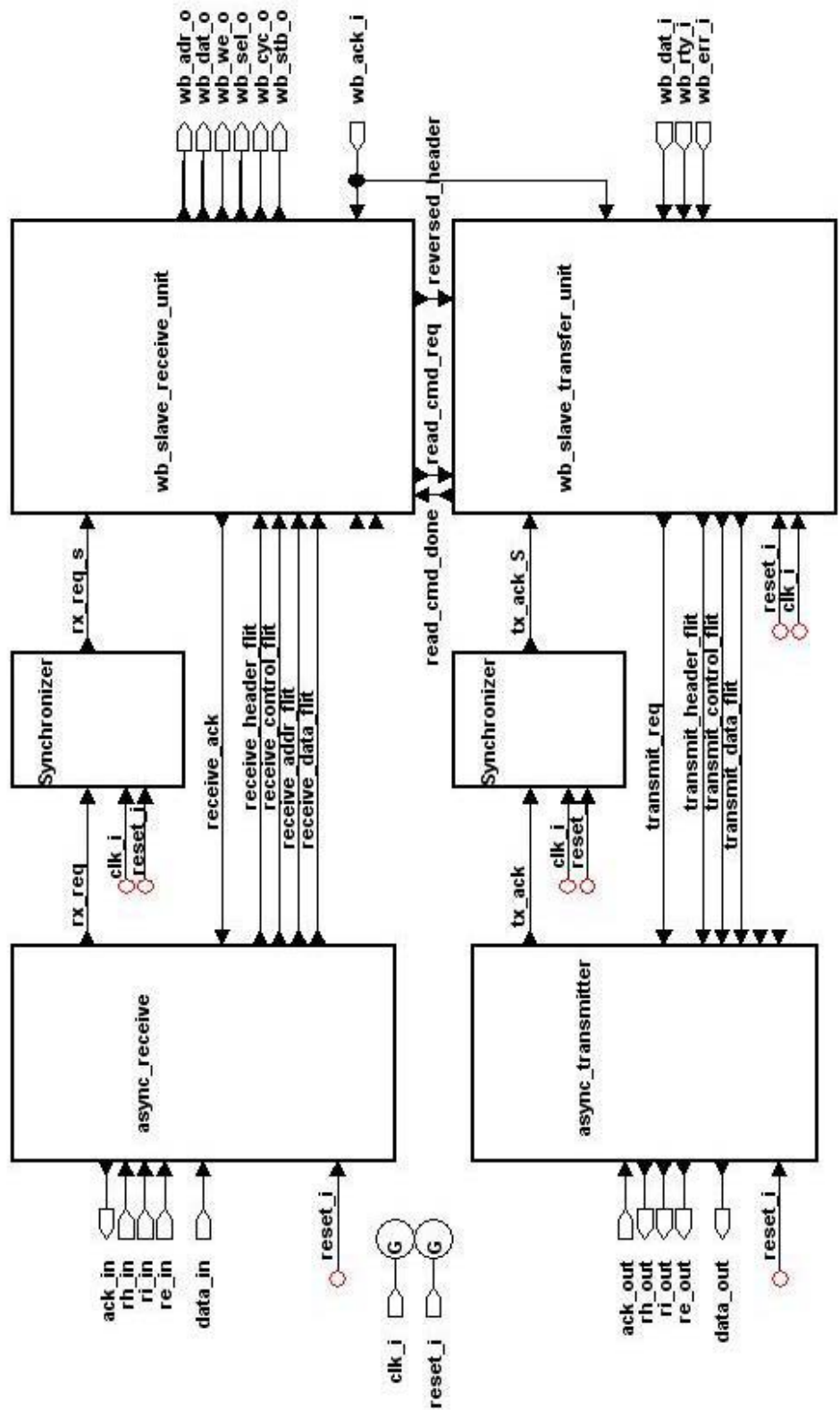

Fig. 6 WISHBONE Slave NA.

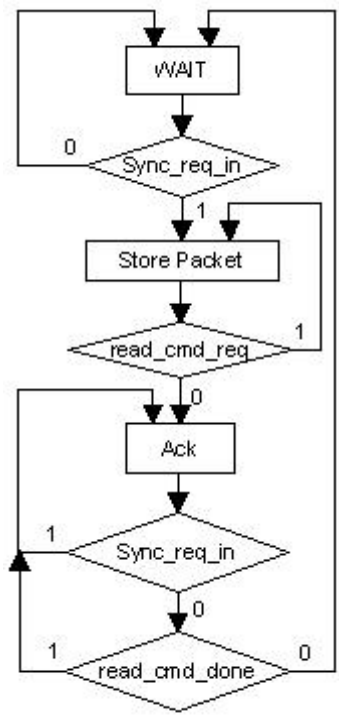

Fig. 5 ASM chart for the WISHBONE master receive unit.

### 4.2 Modified Slave Network Adapters

The Slave NA is very similar to the master counterpart. Figure 6 shows the block diagram for the slave NA. However, the finite state machine for the slave transmit unit and slave receive unit differs.

### 5 Simulation Results

In this work, HDL Designer Pro © - a Mentor Graphics Software [10] - has been used as a graphical design tool where the NoC components have been assembled, configured, and simulated. Also, HDL Designer supports communication with Modelsim © simulator. After successful simulation the SoC is synthesized on a Spartan-3E FPGA device. Table 1 shows the device utilization as reported by Xilinx ISE ver. 11.1. Post-Place & Route static timing report shows that the maximum operating frequency $f_{max}$ = 31.091MHz. The actual frequency of operation was set to 25MHz. This frequency was derived from the 50MHz clock generator available on the Spartan-3E board via a (divide by 2) Digital Clock Manager (DCM) module.

Table 1: Module Level Device Utilization

| Module Name | Slices | Slice Reg | LUTs | LUT RAM |
|---|---|---|---|---|
| Master NA | 98 | 77 | 109 | 0 |
| Slave NA | 136 | 159 | 107 | 0 |

### 6 Conclusions

FPGA devices can be used as an intermediate stage between simulation and ASIC implementations, which reduces verification costs. It has been shown that adding WISHBONE protocol to the core interface side of network adapters enables designers to use the NoC solution in many SoC designs based on the WISHBONE interface. Since, the current implementation support only the basic read and write transaction of the WISHBONE protocol standard, one of the suggested future works is to add support for advanced transactions as well.

**Ahmed H. M. Soliman** has received his B.Sc. degree in Communications, Electronics, and Computer Engineering, Helwan University at 2004. He is currently a tutor at Electronics, Communications, and Computer Engineering Department, Faculty of Engineering, Helwan University, Cairo, Egypt. He is currently interested in Network-on-Chip research issues especially NoC design, prototyping, and parallel processing through NoC connected MPSoC implementations.

**E. M. Saad** has received the B.Sc. degree in electrical (Communication) Eng., Cairo University at 1967, Dipl.-Ing, and Dr-Ing. from Stuttgart University at 1977, 1981 respectively. He is a member of ECS, and EEES. He is currently Prof. of electronic circuits, faculty of Eng. University of Helwan.

**M. El-Bably** received his B.Sc. degree in Communications and Electronics Dept., May 1978, from Helwan University, with grade "Excellent with honor rank". He received M.Sc. degree by courses and dissertation Thesis, from UMIST, Manchester University, U.K, Dec. 1984, Ph.D. degree on testability and diagnosability of digital systems, from Brunel University (west London University), U.K, Feb. 1988.

**Hesham M. A. M. Keshk** has received the B.Sc. degree in Communication and Electronic Engineering Dept., Cairo University at 1982, Master of Science from Helwan University at 1989, Ph.D. from Kyoto University (Japan), at 1996. From 1996 until now he is working in Helwan University. He is interested in Computer Engineering especially parallel processing.